\documentclass[thmsa,11pt]{article}
\usepackage{amsmath}
\usepackage{amssymb}
\usepackage{amsfonts}
\usepackage{graphicx}
\usepackage{color}
\usepackage[stable]{footmisc}
\usepackage{IEEEtrantools}
\usepackage{float}
\usepackage{caption}
\usepackage{subcaption}
\numberwithin{equation}{section}

\restylefloat{figure}

\begin{document}

\title{Optical Properties of Chern-Simons (3+1)D $\theta$-systems}
\author{Luis Huerta$^{1,2}$\\
$^1$  {\small \emph{Departamento de Ciencias Aplicadas, Facultad de Ingenier\'{\i}a, Universidad de Talca.}}\\
$^2$  {\small \emph{P4-Center for Research and Applications in Plasma Physics and Pulsed Power Technology, Chile.}}\\
{\small \texttt{lhuerta@utalca.cl, www.pppp.cl}}}
\maketitle

\begin{abstract}
Chern-Simons (CS) $\theta$-systems are described by a $\theta \int F\wedge F$ term in the action ($\theta$ is an adimensional parameter), which does not change field equations in the bulk, but affects the system behaviour when it is bounded.  When two of those $\theta$-systems, each one characterized by a different value of $\theta$ (even zero), share a common boundary, surface effects are then induced by a CS $\theta$ term. Here, we study the consequences of the above in the propagation of electromagnetic radiation in $\theta$-systems. In a previous paper, electromagnetic radiation properties traversing a $\theta$-vacuum were analyzed where a number of interesting features arise related to polarization and energy distribution. Now, we investigate how electric and magnetic properties of the $\theta$-system affect the optical response. Apart from the well-known Kerr polarization rotation found for the particular case of topological insulators, we examine in detail the issue and the results could be applied in other contexts where $\theta$-term accounts for the system dynamics. In particular, we find two different Brewster angles, for $s$ and $p$ polarization of reflected radiation, respectively, with peculiar features derived from the $\theta$ term influence.  Possible applications of these results are discussed.
\end{abstract}

\section{Introduction}

Chern-Simons (CS) topological theories exhibit a number of remarkable features, which make them interesting as dynamical models for a variety of systems, from gauge field theories to cosmology \cite{Chern-Simons,DJT} (see also \cite{Z-2005}). Moreover, condensed matter physics has benefited from CS approach to give account of nontrivial topology phenomena of certain systems, such as those exhibiting quantum Hall phenomena \cite{QHE}.  In recent years, the novel topological insulators (TI) have become a new example of the above and intensive research has been developed in the area \cite{TI2008,TI2011}. 

In (3+1)D, the CS contribution takes the form of a boundary term, not affecting the behaviour of the system in the bulk.\footnote{This term was introduced in QCD by Roberto Peccei and Helen Quinn in 1977. See \cite{Peccei1977}.}   As it was observed by Frank Wilczek years ago, in spite of the above, the CS $\theta$ term as he called it, could have important applications in describing phenomena of a topological origin \cite{Wilczek87,HZ10}.  In general, the effects of a $\theta$ term in the action for the system are seen when the system is bounded. Then, nontrivial contributions arise from the boundary conditions induced by the CS term, and any physical field reaching the $\theta$-system will be perturbed. An example of this is the modification in the Casimir energy inside a spherical $\theta$-region \cite{Canfora:2011fd}. Effects in gravity are currently under study \cite{CS-Grav}. 

Electromagnetic fields are particularly significant in detecting matter, or even vacuum if the latter is provided by nontrivial properties. When electromagnetic radiation incides on a $\theta$-system, the waves experience changes in their properties, which could be eventually measured. In a previous paper we have pointed out that polarization is affected when reaching the surface of a region characterized as a $\theta$-vacuum. Also, the unexpected presence of a reflected wave in that case subtracts part of the energy crossing the region. We mentioned the possibility of considering this for cosmological issues \cite{HZ12}. In the context of topological insulators, an equivalent phenomenon to Kerr-Faraday polarization rotation occurs in reflected and refracted waves at the surface of such materials \cite{TI2008,TI-Kerr2010}. 

In this report, we describe nonstandard phenomena related to electromagnetic radiation polarization, when incides on a general $\theta$-system, being any kind of matter provided of electric and magnetic properties.  We found that, for some restrictions on the electromagnetic radiation parameters, a nonstandard Brewster angle is possible. These results could be applied for detecting $\theta$-systems and measure their properties.

\section{The basic approach}

If we consider an electromagnetic field in a $\theta$-system, a term of the form
\begin{equation}
\frac{\theta}{2} \int F \wedge F
\end{equation}
is added to the Maxwell action. In noncovariant notation, this amounts to a $\theta\,{\bf E}\cdot{\bf B}$ term (modulus a constant). Then, the field equations become
\begin{eqnarray}
\varepsilon\mathbf{\nabla }\cdot \mathbf{E} &\mathbf{=}&\theta \delta (\Sigma )\mathbf{B}\cdot \mathbf{n}
 \label{Gauss} \\
\frac{1}{\mu}\nabla \times \mathbf{B} -\partial_{t}\mathbf{E}&\mathbf{=}&\theta \delta (\Sigma) \mathbf{E\times n} .
 \label{Ampere}
\end{eqnarray}
The delta function, $\delta(\Sigma)$, means that RHS of the above equations are evaluated at the interface separating the system for a surrounding normal medium ($\mathbf{n}$ is the unit normal to $\Sigma $). Electric permittivity and magnetic permeability are added to consider a more general situation.\footnote{We use $c=1$ units.} What we see is that the interface partially transforms electric and magnetic fields into each other, and represent surface charge and current densities, respectively \cite{HZ12}.  

In the vicinity of the surface $\Sigma$, these equations imply that the normal component of $\mathbf{E}$ and the tangential component of $\mathbf{B}$ are discontinuous:
\begin{eqnarray}
\left[ \varepsilon {\bf E}_{\rm n} \right] &=& \left. \theta {\bf{B}}_{\rm n}\right|_{\Sigma} \label{Emat_discon} \\
\left[ \dfrac{1}{\mu} {\bf B}_\tau \right]  &=& \left. - \theta {\bf E}_\tau \right|_{\Sigma} . \label{Bmat_discon} 
\end{eqnarray}
Subindices $n$ and $\tau$ stand for normal and tangent components of the fields, respectively, with respect to the interface $\Sigma$. The symbol $\left[ {\,\,} \right]$ must be interpreted as the difference between the fields evaluated immediately inside and immediately outside the $\theta$-medium.  In addition to those boundary discontinuities, we have the boundary conditions derived from the identity $dF\equiv 0$, thus leading to the standard continuity conditions for the tangential component of $\mathbf{E}$ and the normal component of $\mathbf{B}$, at the interface $\Sigma$,
\begin{eqnarray}
\left[ {\bf{E}}_{\tau} \right] = 0  \label{Evac_con} \\
\left[ {\bf{B}}_n \right] = 0 .  \label{Bvac_con}
\end{eqnarray} 
Notice that these latter continuity conditions ensure consistency of (\ref{Emat_discon}) and (\ref{Bmat_discon}).

\section{Optical properties of $\theta$-systems}

Consider a small region in the boundary between the two media where an electromagnetic wave incides. The wave comes from a non-$\theta$ surrounding region and incides the system with an angle $\varphi$.  The wave is reflected with the same angle $\varphi$ and transmitted to the other medium with a different angle $\psi$, all these angles measured respect to the normal to the interface as usual.\footnote{The usual laws of reflection and refraction hold, independently of the wave polarization state. This is also true under the presence of a CS term in the action.} We adopt the standard decomposition of the electric and magnetic fields in a $p$ (parallel) and an $s$  (perpendicular) component, with respect to the incidence plane. 

Also the electric and magnetic fields in each medium are related by the standard relationship ${\mathbf{B}} = n{\bf{\hat k}} \times {\mathbf{E}} \label{E_B}$, where $n = {\left( \mu \varepsilon \right)^{1/2}}$ is the respective refraction index of the system and $\hat{\bf k}$ is the corresponding unit wave vector. This relationship permits us to express results only in terms of the electric field. We introduce the convenient definitions ${e_{i\parallel }} \equiv {E_{i\parallel }}/{E_i}$, ${e_{r\parallel }} \equiv {E_{r\parallel }}/{E_i}$, ${e_{t\parallel }} \equiv {E_{t\parallel }}/{E_i}$, with ${E_i} = {\left( {{E_{i\parallel }}^2 + {E_{i \bot }}^2} \right)^{1/2}}$, and the same for $s$ components. By applying the b.~c. (\ref{Evac_con})$-$(\ref{Bvac_con}) and (\ref{Emat_discon})$-$(\ref{Bmat_discon}), we find for the refracted wave (subscript $t$ stands for transmitted),
\begin{equation}
\left( {\begin{array}{*{20}{c}}
{{e_{t\parallel }}}\\
{{e_{t \bot }}}
\end{array}} \right) = \frac{2}{D}\left( {\begin{array}{*{20}{c}}
{\eta s + 1}&{ - \theta }\\
{\theta s}&{\eta  + s}
\end{array}} \right)\left( {\begin{array}{*{20}{c}}
{{e_{i\parallel }}}\\
{{e_{i \bot }}}
\end{array}} \right) \label{e_t}
\end{equation}
and for the reflected wave,

\begin{equation}
\left( {\begin{array}{*{20}{c}}
{{e_{r\parallel }}}\\
{{e_{r \bot }}}
\end{array}} \right) = \frac{1}{D}\left( {\begin{array}{*{20}{c}}
{\left( {\eta s + 1} \right)\left( {\eta  - s} \right) + {\theta ^2}s}&{2\theta s}\\
{2\theta s}&{ - \left( {\eta s - 1} \right)\left( {\eta  + s} \right) - {\theta ^2}s}
\end{array}} \right)\left( {\begin{array}{*{20}{c}}
{{e_{i\parallel }}}\\
{{e_{i \bot }}}
\end{array}} \right) \label{e_r}
\end{equation}
where $D \equiv \left( {\eta s + 1} \right)\left( {\eta  + s} \right) + { \theta^2}s$; $s \equiv \cos \psi /\cos \varphi$, where $\varphi$ is the angle of incidence and $\psi$, the angle of refraction; ${\eta _{1(2)}} \equiv {n_{1(2)}}/\mu_{1(2)}$, and $\eta  \equiv {\eta _2}/{\eta _1}$, for the dielectric and magnetic properties ($\mu$ is the magnetic permeability of the respective medium and $n$, its index of refraction). To simplify notation, we have redefined $\theta \to \theta /{\eta _1}$.\footnote{Of course, if the surrounding medium is vacuum space, $\theta$ coincides with the parameter in the action.} 

\begin{figure}[H]
        \centering
        \begin{subfigure}[b]{0.5\textwidth}
                \includegraphics[width=\textwidth]{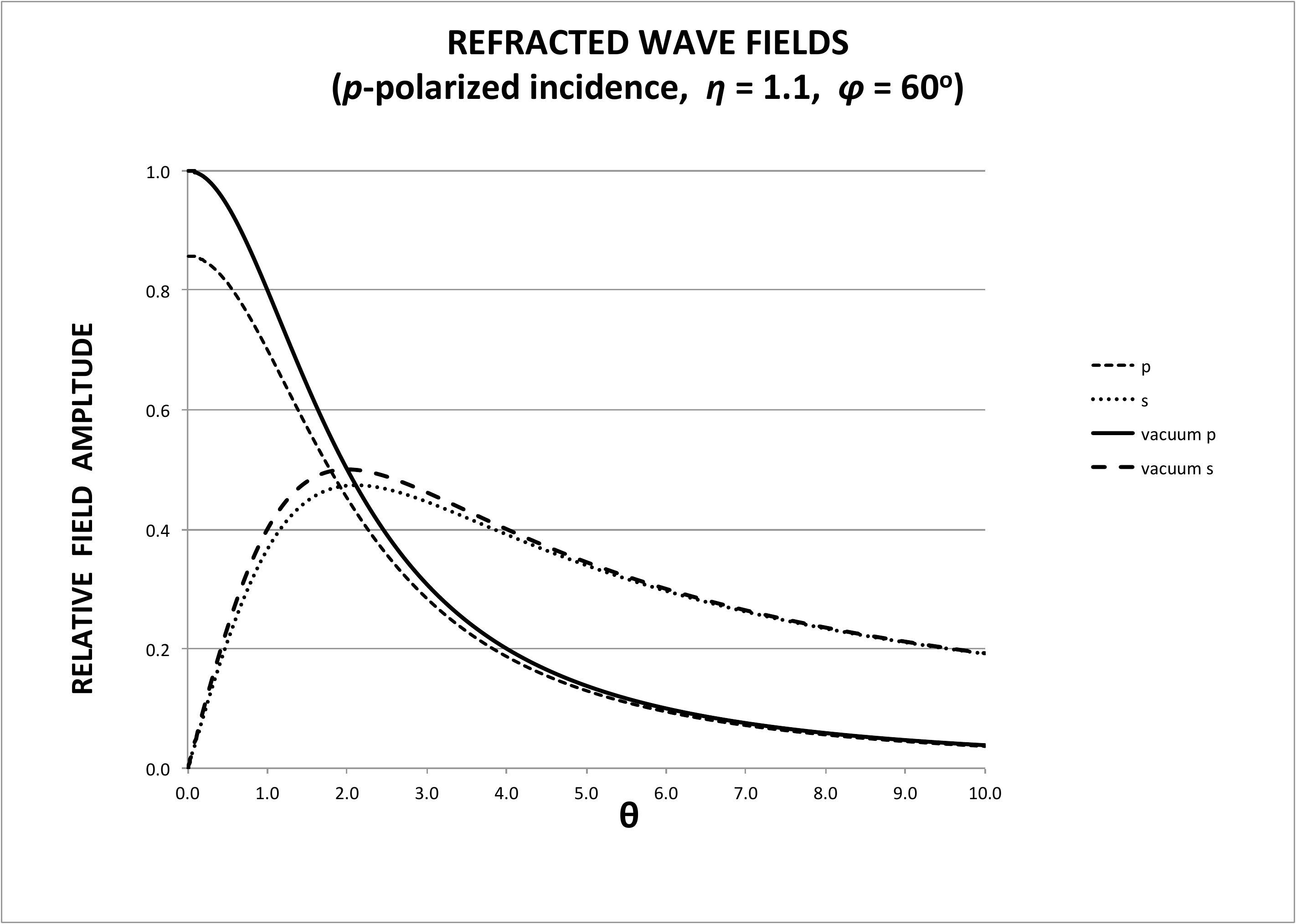}
                \caption{}
                \label{fig_tFields}
        \end{subfigure}%
        ~ 
        \begin{subfigure}[b]{0.5\textwidth}
                \includegraphics[width=\textwidth]{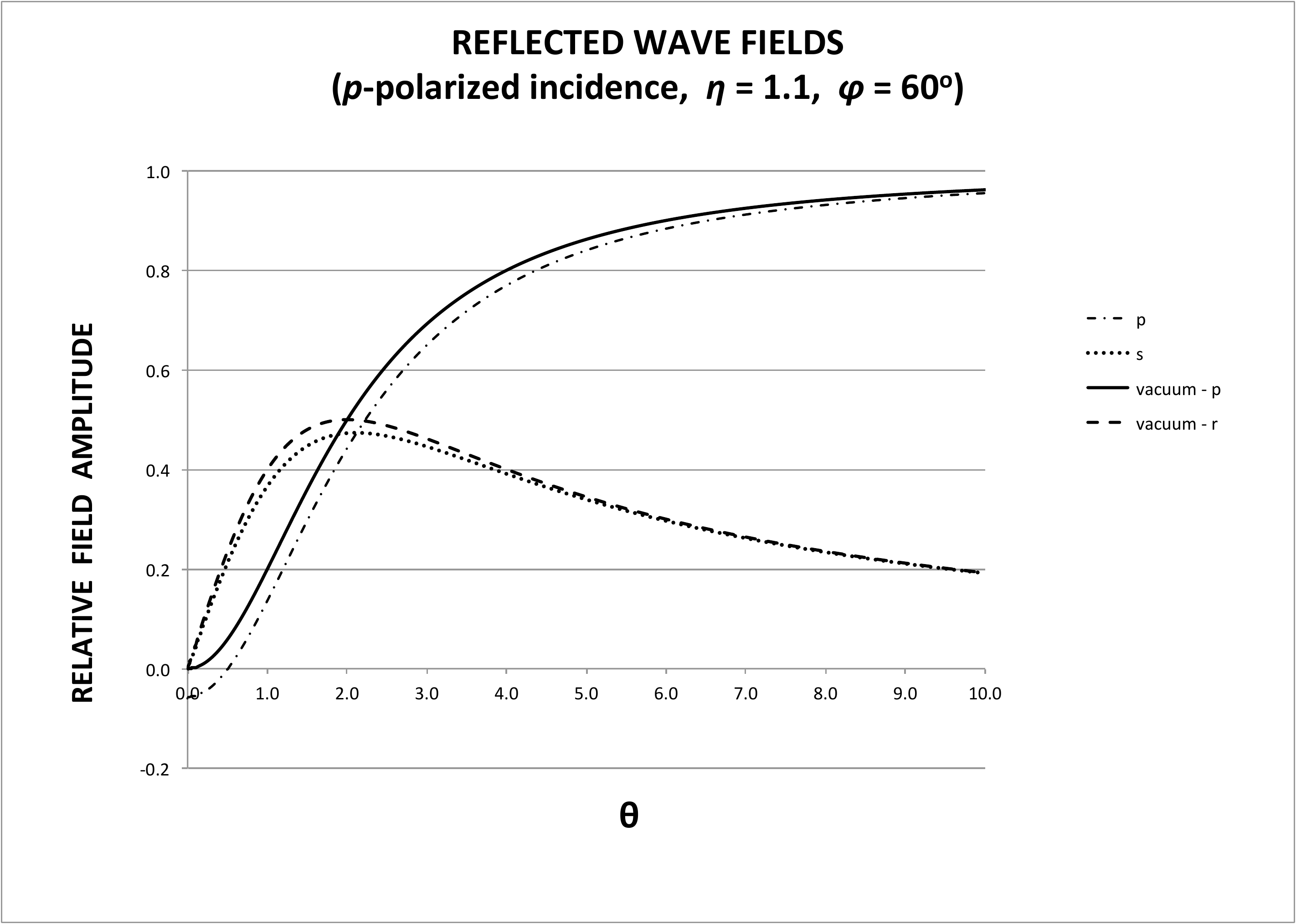}
                \caption{}
                \label{fig_rFields}
        \end{subfigure}
        ~ 
               \caption{Field amplitudes for refracted and reflected waves. Comparison is made with the $\theta$-vacuum case. Notice the value of $\theta$ for the Brewster angle ($e_{r\parallel} =0$) in the reflected wave.}
               \label{fig_Fields}
\end{figure}

Figure \ref{fig_Fields} shows the amplitude fields both for refracted and reflected waves compared to the $\theta$-vacuum.  For the chosen value of $\eta=1.1$ (and $\phi=60^o$), we find that the reflected wave presents a Brewster angle for $\theta=0.504$. This issue will be discussed in more detail below.

\vspace{0.5cm}

\section{Polarization rotation}

Unlike the normal non-$\theta$ systems, the $p$ and $s$ components of fields mix with each other for both the reflected and refracted waves, and those components do not vanish even when the corresponding component of the incident wave is not originally present. Both waves experiment a polarization plane rotation, the well-known Kerr-Faraday rotation which occurs when light interacts with a magnetic material (Kerr case corresponds to the reflection of light by a surface) \cite{Kerr}. The polarization angle, $\alpha$, is defined by $\tan \alpha  = e_ \bot / e_\parallel$. The refracted wave polarization angle is given by
\begin{equation}
{\alpha _t} = \tan^{-1} \left[ \frac{{\theta s + \left( {\eta  + s} \right)\tan {\alpha _i}}}{{\eta s + 1 - \theta \tan {\alpha _i}}}\right] \label{tPol}
\end{equation}
and, for the reflected wave,
\begin{equation}
{\alpha _r} = \tan^{-1} \left[ \frac{{2\theta s - \left[ {\left( {\eta s - 1} \right)\left( {\eta  + s} \right) + {\theta ^2}s} \right]\tan {\alpha _i}}}{{\left( {\eta s + 1} \right)\left( {\eta  - s} \right) + {\theta ^2}s + 2\theta s\tan {\alpha _i}}} \right] . \label{rPol}
\end{equation}
$\alpha_i$ represents the incident wave polarization angle. 

\begin{figure}[H]
        \centering
        \begin{subfigure}[b]{0.5\textwidth}
                \includegraphics[width=\textwidth]{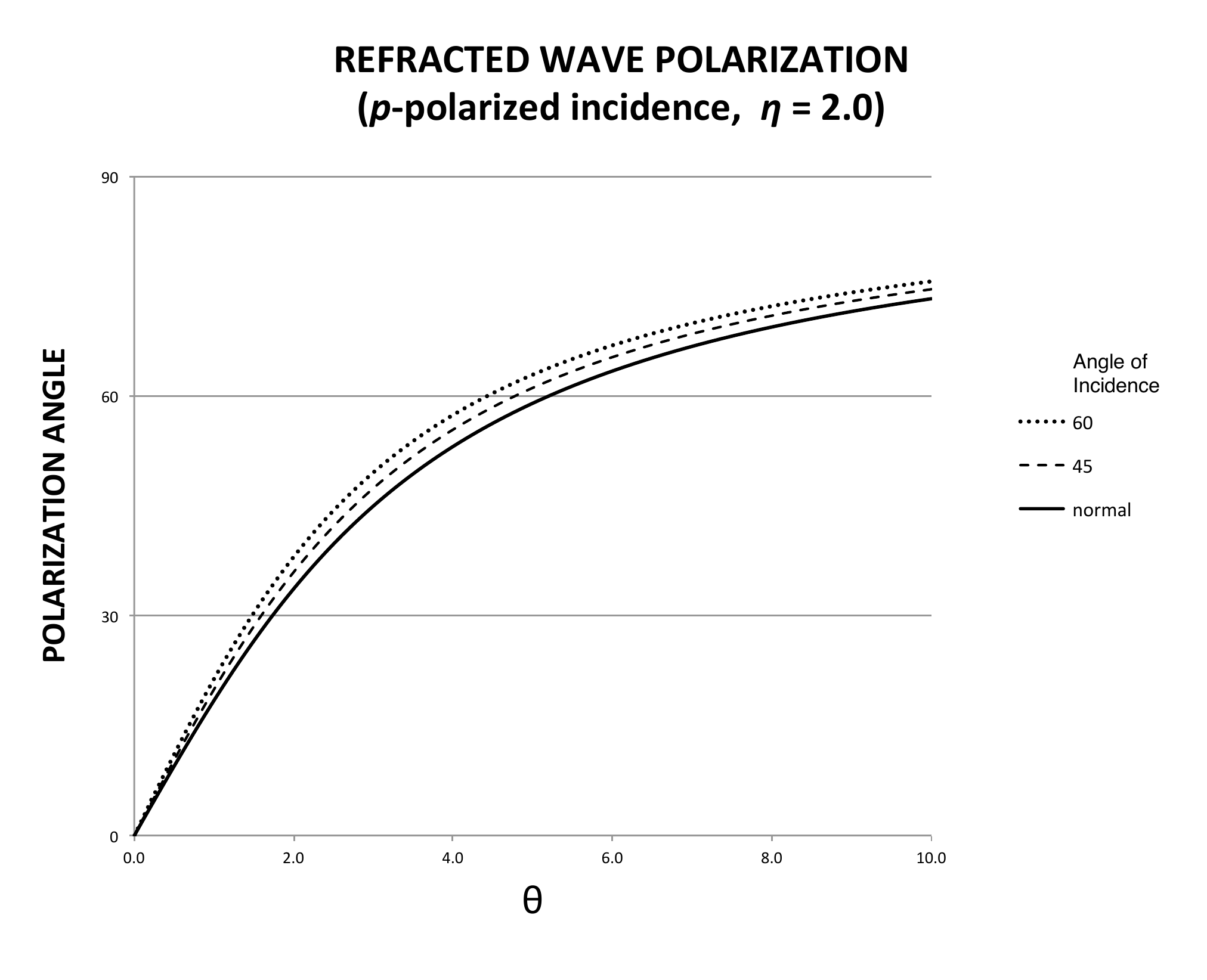}
                \caption{}
                \label{fig_tPol}
        \end{subfigure}%
        ~ 
        \begin{subfigure}[b]{0.5\textwidth}
                \includegraphics[width=\textwidth]{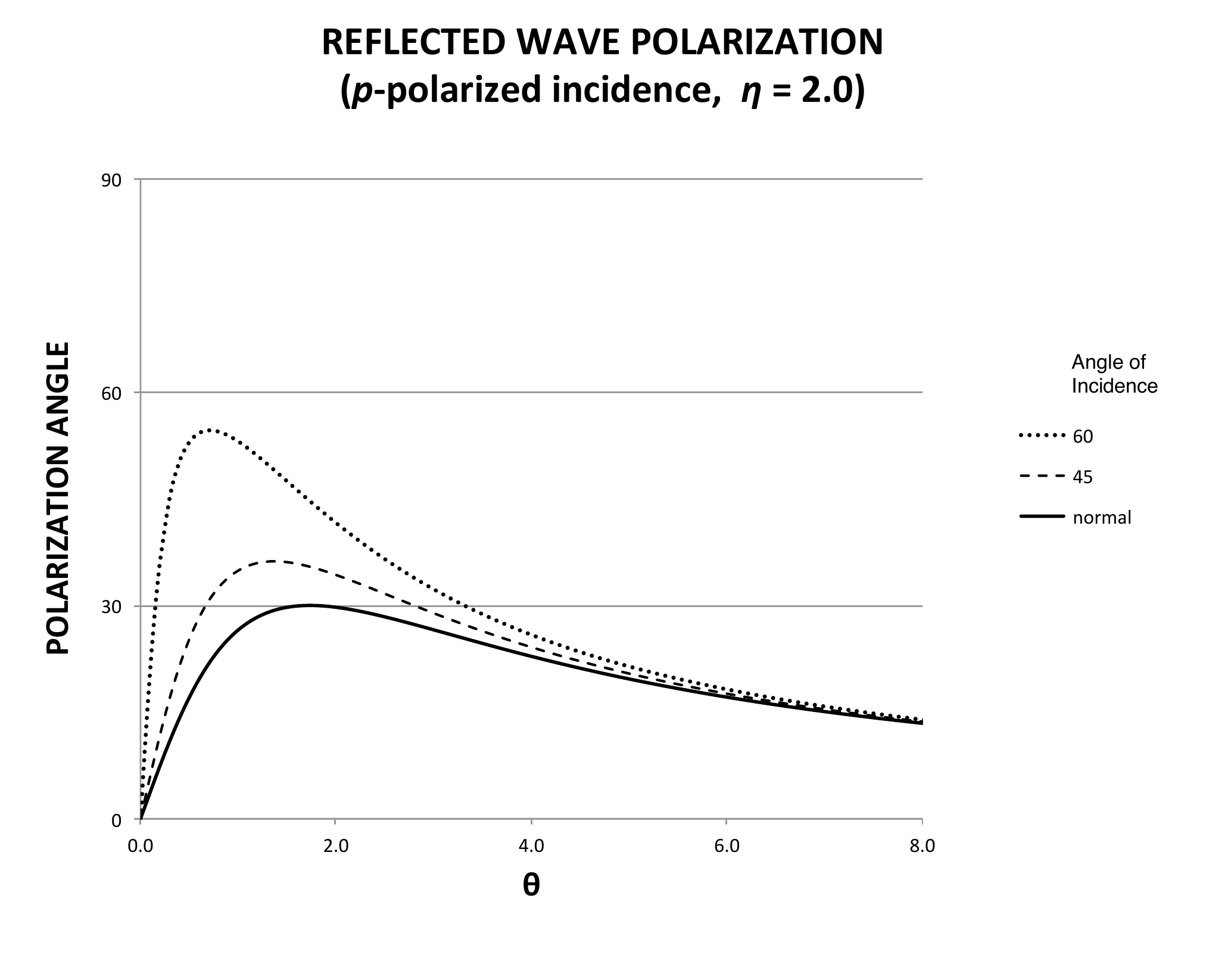}
                \caption{}
                \label{fig_rPol}
        \end{subfigure}
        ~ 
               \caption{Refracted and reflected wave polarization for different incidence angles.}
               \label{fig_Pol}
\end{figure}

Figure \ref{fig_Pol} shows the polarization angle for different values of $\theta$ in the case of a $p$-polarized ($\alpha_i =0$) incident wave incoming at different angles of incidence. Because of such a polarization, the curves actually represent the rotation in polarization experienced by the outgoing waves. We see that the $\theta$ term has a significant influence and, in the case of reflected waves, we find a maximum polarization rotation for $\theta _{\max }^{\left( r \right)} = \sqrt {\left( {\eta  + \frac{1}{s}} \right)\left( {\eta  - s} \right)}$, with a value for the polarization given by
\begin{equation}
{\alpha _{r,\,\max }} = {\tan ^{ - 1}}{\left[ {\left( {\eta  + \frac{1}{s}} \right)\left( {\eta  - s} \right)} \right]^{ - 1/2}} . \label{rPol_max}
\end{equation}
This maximum depends on both incidence angle and the system properties (beyond $\theta$ dependence) given by $\eta$ parameter.

\vspace{0.5cm}

\vspace{0.5cm}
\begin{figure}[H]
        \centering
        \begin{subfigure}[b]{0.5\textwidth}
                \includegraphics[width=\textwidth]{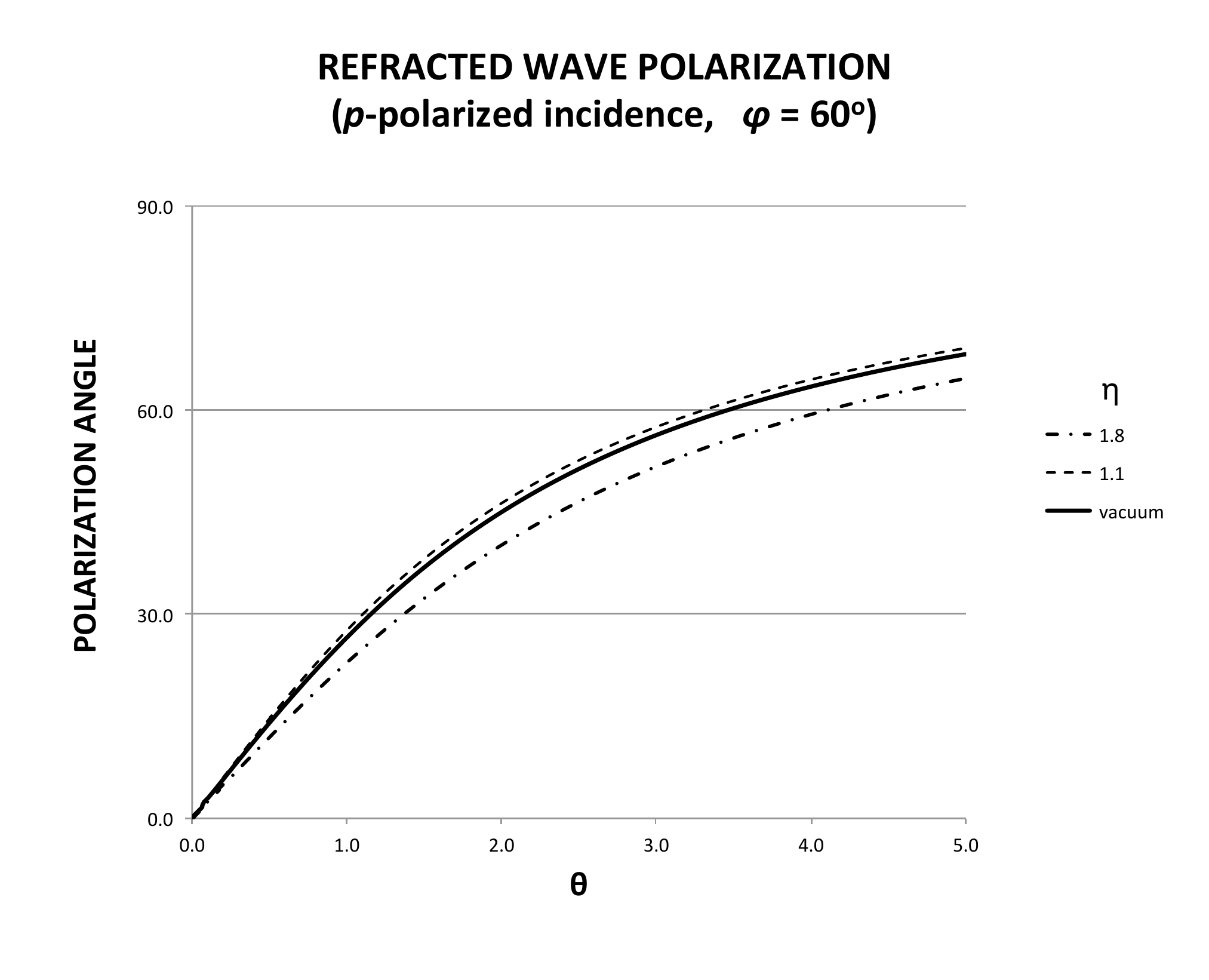}
                \caption{}
                \label{fig_tPol-eta}
        \end{subfigure}%
        ~ 
        \begin{subfigure}[b]{0.5\textwidth}
                \includegraphics[width=\textwidth]{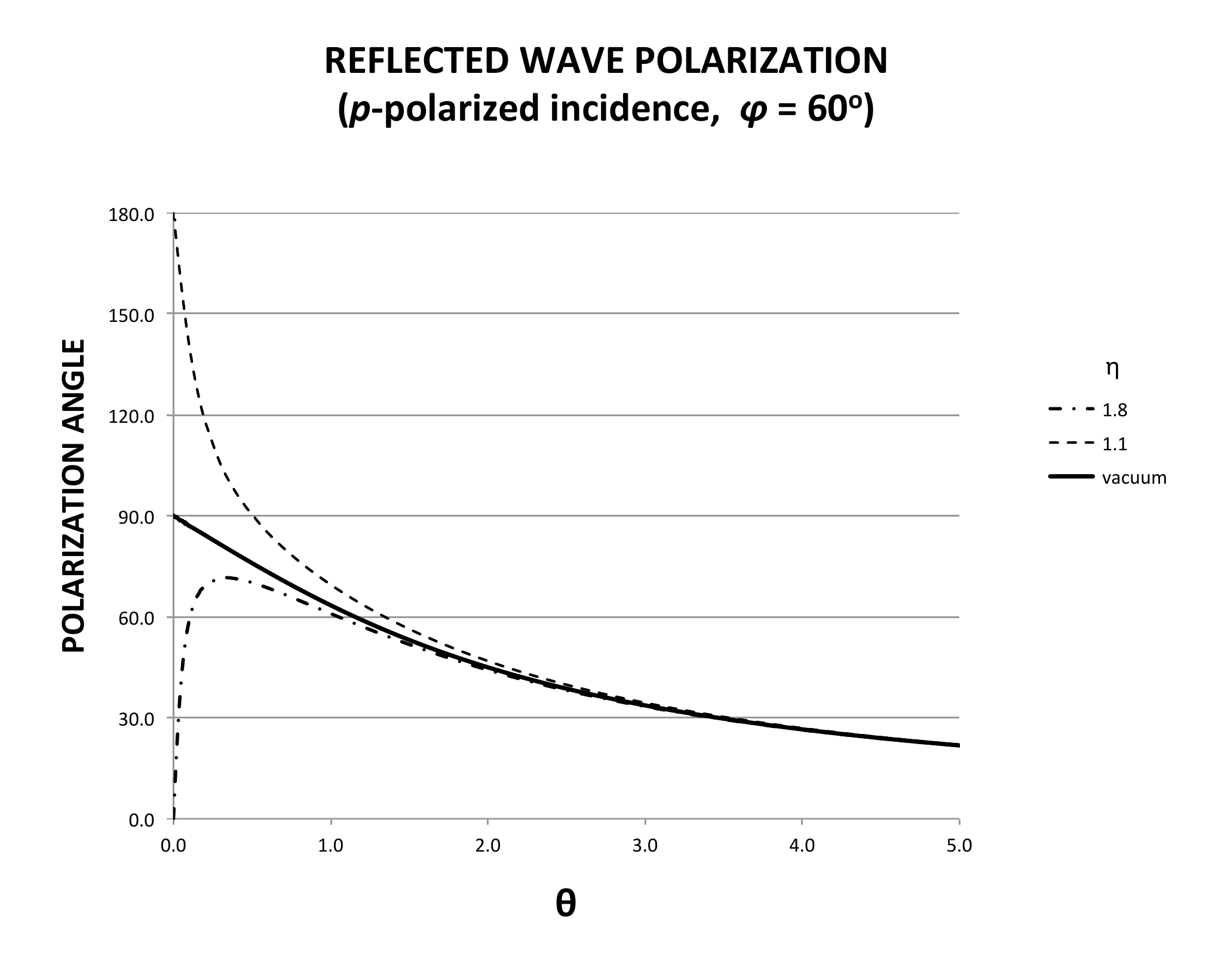}
                \caption{}
                \label{fig_rPol-eta}
        \end{subfigure}
        ~ 
               \caption{Refracted and reflected wave polarization for different system electric properties (we choose $\mu$=1 here).}
               \label{fig_Pol-eta}
\end{figure}

In Fig. \ref{fig_Pol-eta} we show the refracted and reflected waves waves polarization rotation compared to $\theta$-vacuum case. For the reflected waves we see a different behaviour for the lower value of $\eta=1.1$ and the greater value of $\eta=1.8$, with the $\theta$-vacuum case in the middle of them. This is so because, the numerator in (\ref{rPol}) changes its sign for $\eta=\tan\phi$, that is, for the Brewster angle condition for normal non-$\theta$ systems. For the value of $\phi=60^o$, chosen for the curves in the figure, the latter amounts to $\eta=\sqrt(3)$. 

\vspace{0.5cm}
\begin{figure}[H]
        \centering
        \begin{subfigure}[b]{0.5\textwidth}
                \includegraphics[width=\textwidth]{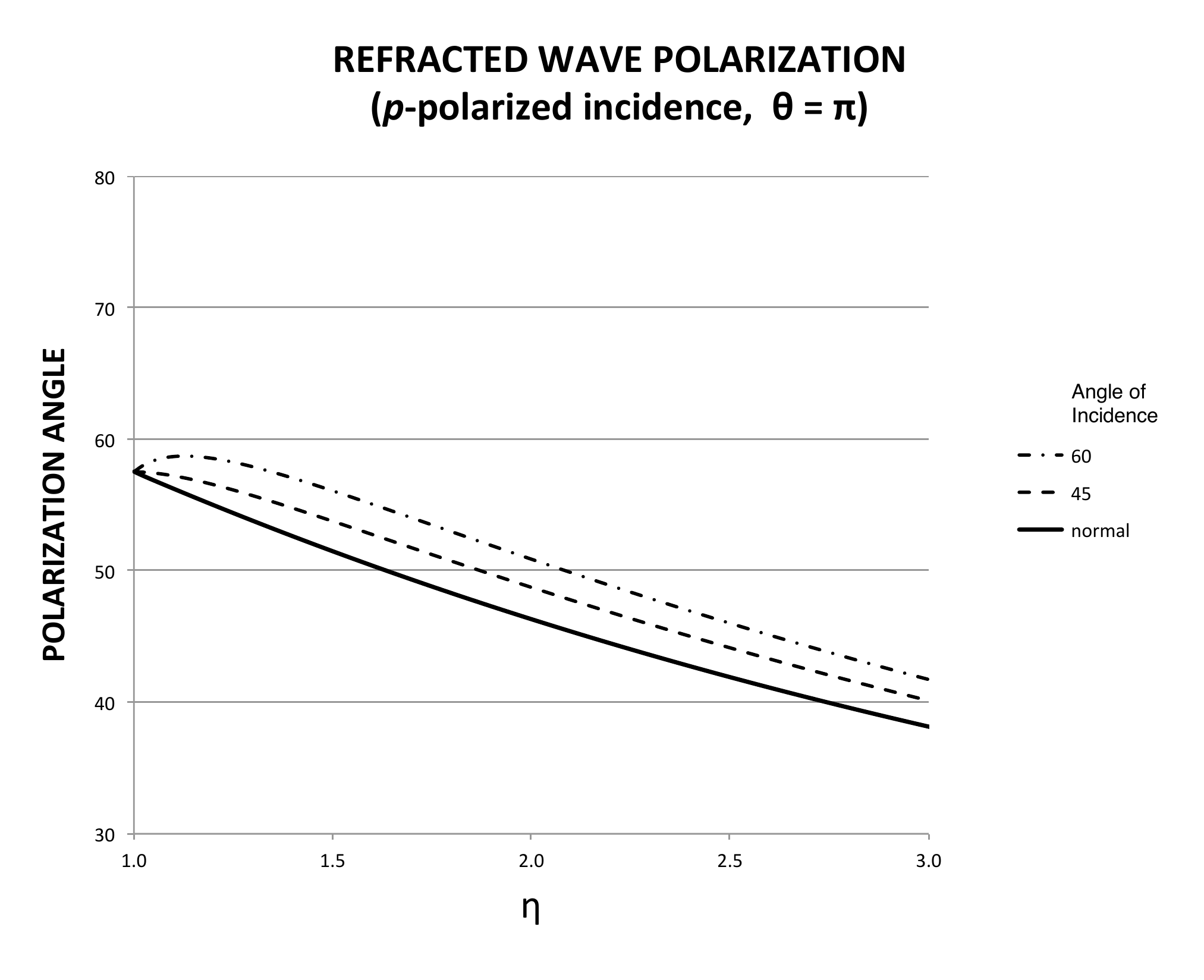}
                \caption{}
                \label{fig_tPol_eta}
        \end{subfigure}%
        ~ 
        \begin{subfigure}[b]{0.5\textwidth}
                \includegraphics[width=\textwidth]{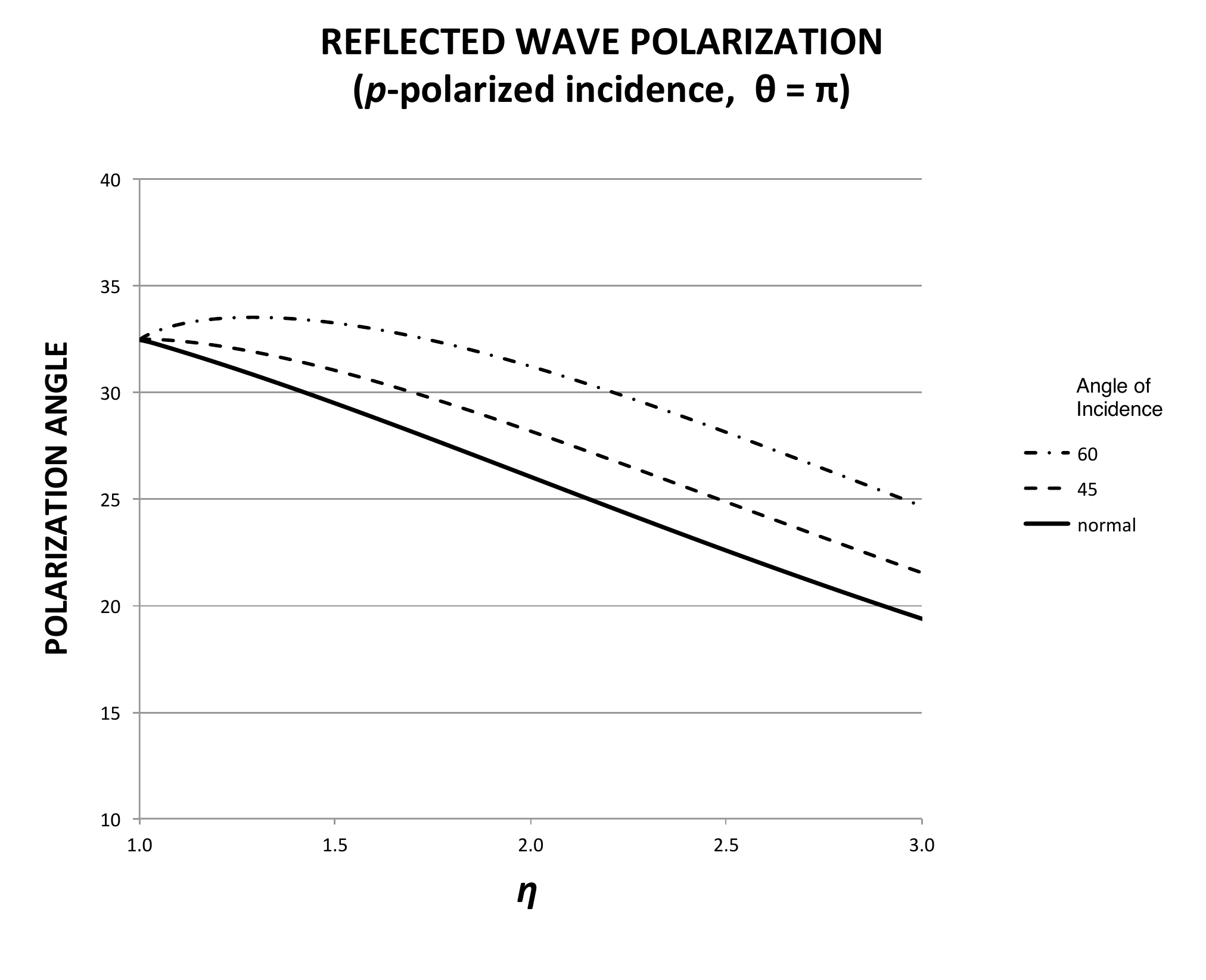}
                \caption{}
                \label{fig_rPol_eta}
        \end{subfigure}
        ~ 
               \caption{Refracted and reflected wave polarization at $\theta$ interface: dependence on electric and magnetic system properties (represented by the parameter $\eta$).}
               \label{fig_Pol_eta}
\end{figure}

The dependence on $\eta$ is shown in Fig. \ref{fig_Pol_eta}. We observe that the possible electric and magnetic properties diminish the influence of $\theta$, and polarization diminishes as $\eta$ grows.  Maximum polarization rotation is exhibited by a $\theta$-vacuum. However, a remarkable feature arises since, above certain value of the incidence angle, a maximum is possible ffor an $\eta > 1$, both for refracted and reflected waves. A general expression for the value of $\eta$ for this polarization maximum, at fixed incidence angle, is rather cumbersome to write. However, it can be demonstrated that the condition for the incidence angle in order to see the phenomenon is $\phi>45^o$. For instance, for an incidence angle of $\phi=60^o$, we find that the maximum is obtained for $n =1.29$ for the reflected wave, and for $n =1.13$ in the case of the refracted wave (we consider nonmagnetic systems, thus $\eta=n$).

\section{Brewster angle for $\theta$-systems}

The radiation reflected by a nonconducting surface could eventually become $s$-polarized; that is, the reflected wave polarization plane lies perpendicular to the incidence plane (and, consequently, parallel to the interface surface). This occurs for a particular angle of incidence, which we call here the $s$-Brewster angle, for a reason that will become apparent immediately. This phenomenon also occurs for electromagnetic radiation inciding on $\theta$-systems. But, the novelty is that a nonstandard Brewster angle appears, where reflected waves become polarized in the plane of incidence.   We call the latter angle the $p$-Brewster angle. Let us first consider the standard $s$-Brewster angle. Then, by putting $e_r=0$ in (\ref{e_r}), an equation for $s$ is obtained. Since the parameter $s$ is related to the incidence angle by $\tan \varphi  = n\sqrt{(s^2 - 1)/(n^2- 1)}$, where $n$ is the refraction index, we find
\begin{equation}
\tan{\varphi ^{(s)}_B} = \frac{1}{2\sqrt {\eta ^2 - 1/\mu^2}} \left\{ \left[ B(\theta)+2\theta\tan\alpha_i+ \sqrt{ \left( B(\theta)+2\theta\tan\alpha_i \right)^2 + 4\eta ^2} \right]^2 - 4\eta ^2 \right\}^{1/2}  
\label{Brewster} ,
\end{equation}
where we have introduced, for convenience, $B(\theta) \equiv \eta ^2 - 1 + \theta ^2$.  

\begin{figure}[H]
        \centering
        \begin{subfigure}[b]{0.5\textwidth}
                \includegraphics[width=\textwidth]{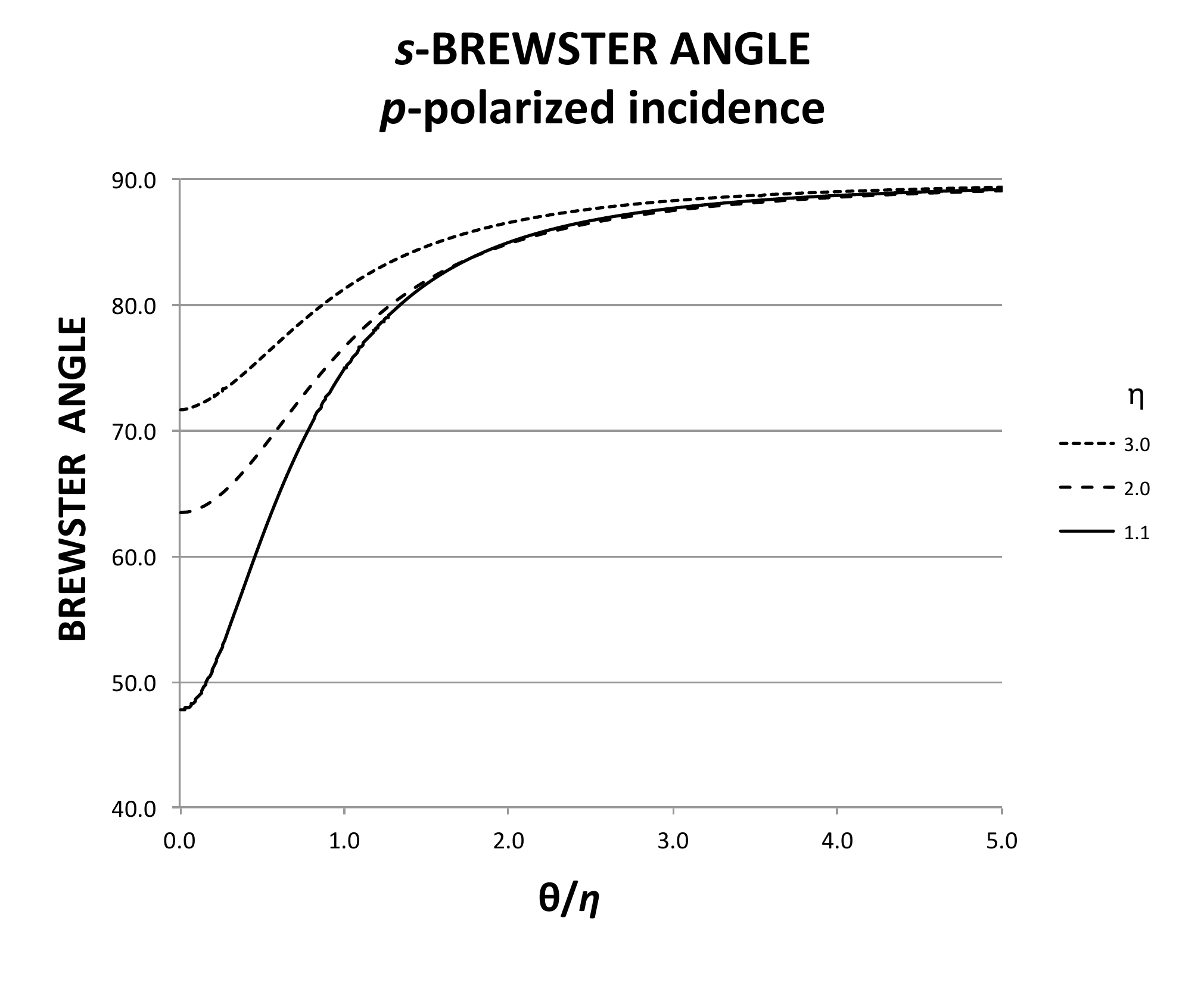}
                \caption{}
                \label{fig_Brewster_a}
        \end{subfigure}%
        ~ 
        \begin{subfigure}[b]{0.5\textwidth}
                \includegraphics[width=\textwidth]{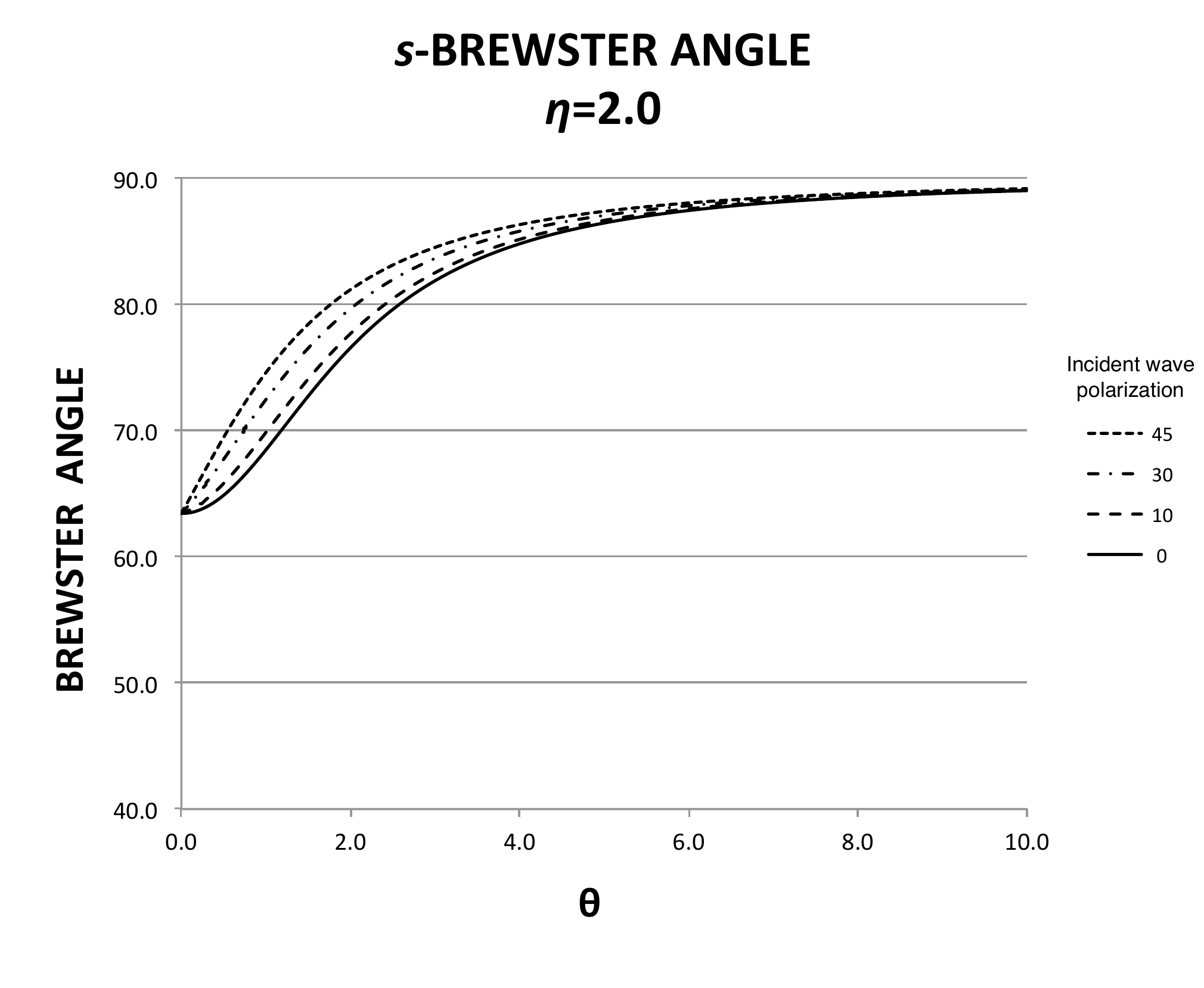}
                \caption{}
                \label{fig_Brewster_b}
        \end{subfigure}
        ~ 
               \caption{Standard $s$-Brewster angle is shown in terms of $\theta$. (a) The curves are computed for different values of $\eta$ and when incident wave is $p$-polarized. The abscissa was chosen as the quotient $\theta/\eta$, for clarity. (b) Different curves represent different values of the incident wave polarization angle $\alpha_i$. Dielectric and magnetic properties of the system are given by $\eta=2.0$.}
               \label{fig_Brewster}
\end{figure}

The results are plotted in Fig.~\ref{fig_Brewster}. Several interesting points arise. Since the $\theta$ term in the equations has a significant influence in polarization, this is also seen in the values for Brewster angle. Even for rather small values of $\theta$, the standard $s$-Brewster angle is changed notoriously [see Fig.~\ref{fig_Brewster_a}]. On the other hand, large values of $\theta$, compared to $\eta$, cancel the possibility of having a $s$-Brewster angle, since the value of the angle tends to $90^o$. Additionally, an important difference with normal media is that the standard $s$-Brewster angle depends on incident wave polarization, $\alpha_i$. The greater the values of this polarization angle, the lower the values of $\theta$ sufficient for changing significantly the Brewster angle.\footnote{The extreme case, $\alpha_i =90^o$, that is, when the incident wave electric field has no component parallel to the plane of incidence ($s$-polarized incidence), illustrates this.}

\begin{figure}[H]
        \centering
        \begin{subfigure}[b]{0.5\textwidth}
                \includegraphics[width=\textwidth]{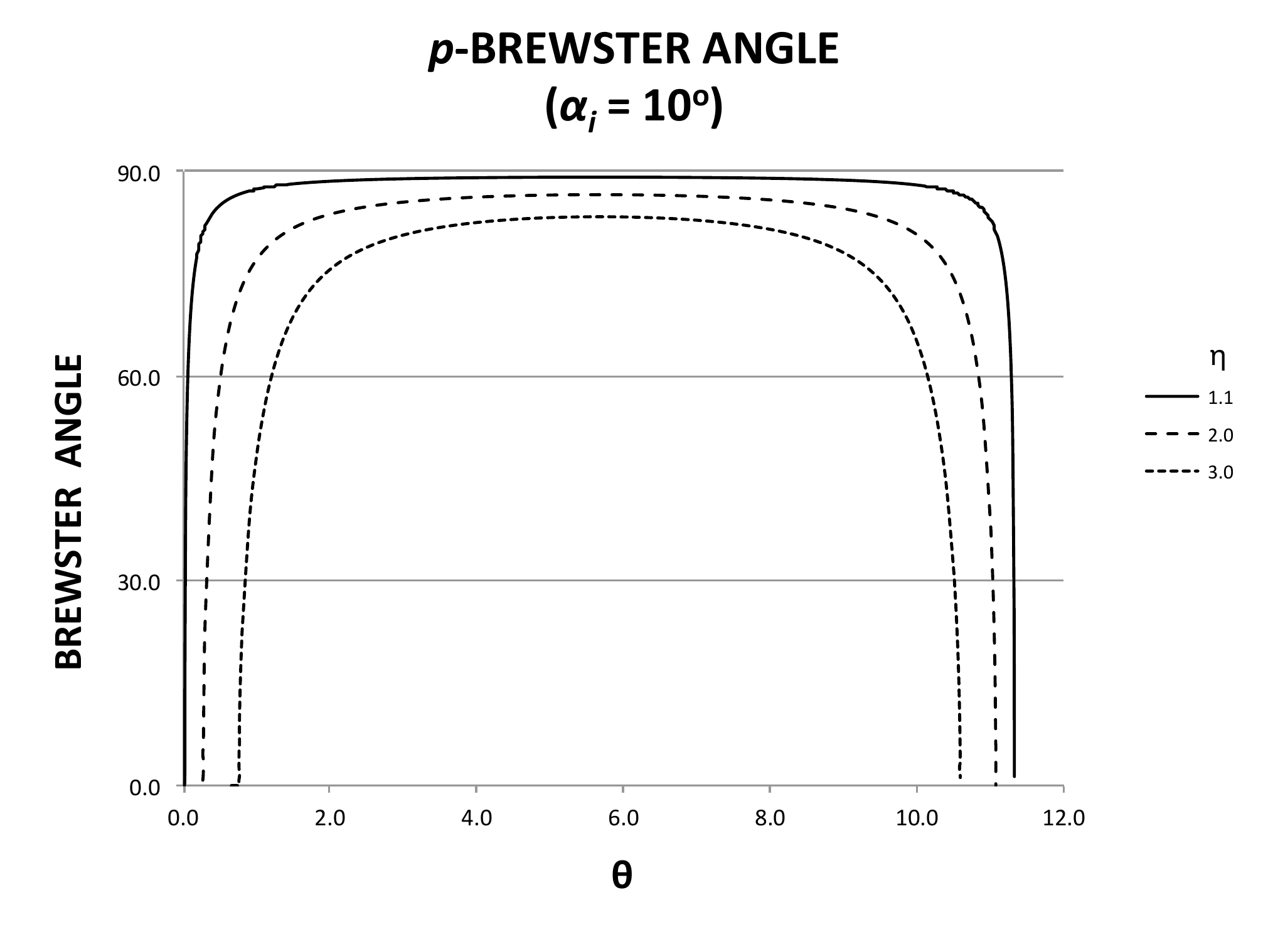}
                \caption{}
                \label{}
        \end{subfigure}%
        ~ 
        \begin{subfigure}[b]{0.5\textwidth}
                \includegraphics[width=\textwidth]{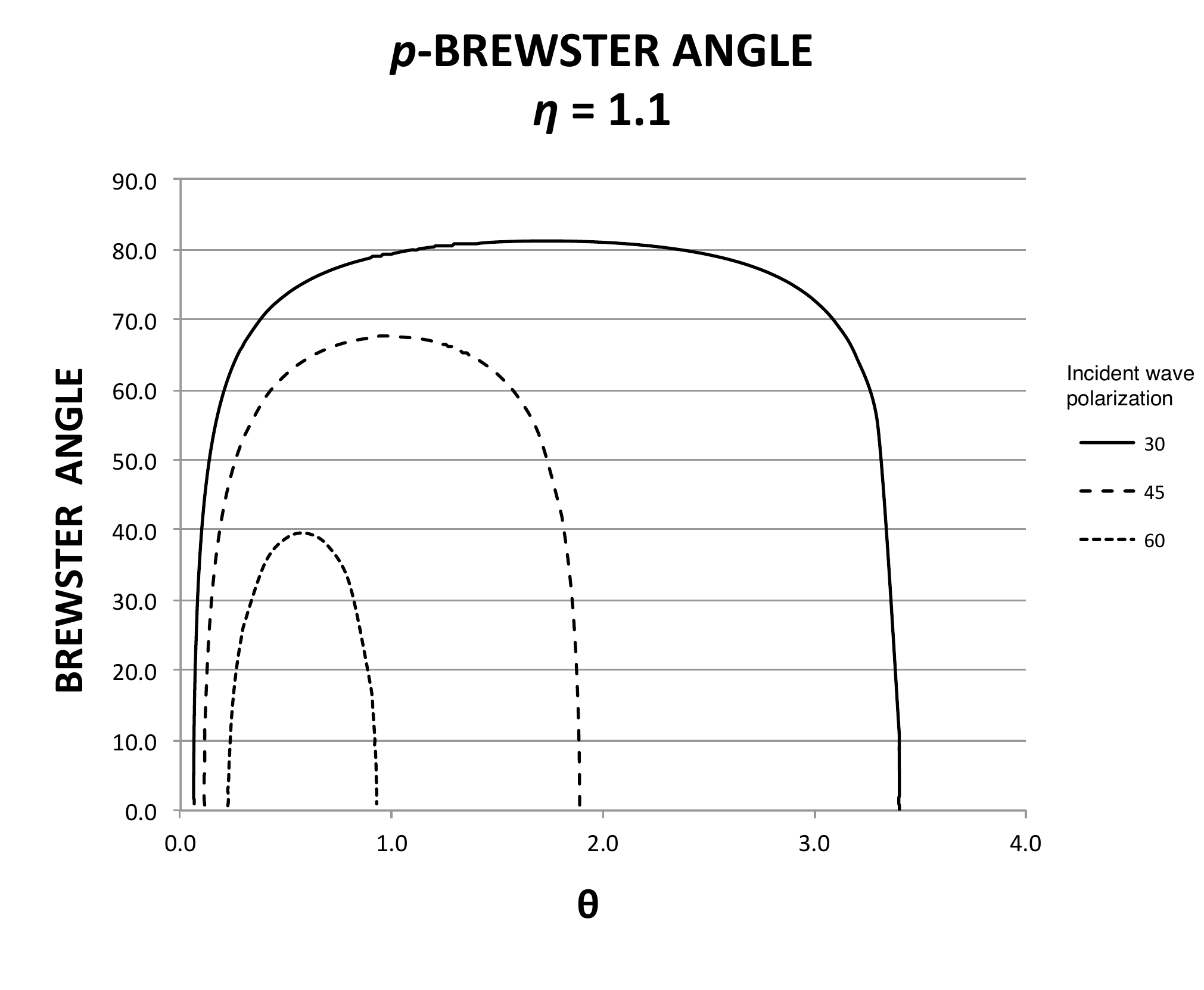}
                \caption{}
                \label{}
        \end{subfigure}
        ~ 
               \caption{Nonstandard $p$-Brewster angle versus $\theta$. (a) The curves are for different values of $\eta$, and the incident wave polarization angle is $\alpha_i = 10^o$. (b) $p$-Brewster angle for different values of the incident wave polarization angle, $\alpha_i$. Dielectric and magnetic properties of the system are represented by $\eta = 1.1$.}\label{fig_nonBrewster}
\end{figure}

The nonstandard $p$-Brewster angle is obtained by imposing  $e_{r\perp} = 0$ in (\ref{e_r}), thus yielding
\begin{equation}
\varphi _B^{\left( p \right)} = {\tan ^{ - 1}}\left\{ {\frac{1}{{2\sqrt {{\eta ^2} - 1/{\mu ^2}} }}{{\left[ {{{\left( { - B\left( \theta  \right) + 2\cot {\alpha _i} + \sqrt {{{\left( {B\left( \theta  \right) - 2\cot {\alpha _i}} \right)}^2} + 4{\eta ^2}} } \right)}^2} - 4{\eta ^2}} \right]}^{1/2}}} \right\} .
\label{non_Brewster} 
\end{equation}

We see that the $p$-Brewster angle does exist only for a restricted set of values for $\theta$, namely,
\begin{equation}
\cot {\alpha _i} - \sqrt {{{\cot }^2}{\alpha _i} - {\eta ^2} + 1}  < \theta  < \cot {\alpha _i} + \sqrt {{{\cot }^2}{\alpha _i} - {\eta ^2} + 1} . \label{nonB_notheta}
\end{equation}
In addition, the latter set for $\theta$ is not empty if $\alpha_i$, the incident wave polarization, satisfies the inequality
\begin{equation}
{\alpha _i} < {\tan ^{ - 1}}\frac{1}{{\sqrt {{\eta ^2} - 1} }} .  \label{nonB_nopol}
\end{equation}

Figure \ref{fig_nonBrewster} shows the $p$-Brewster angles for different values of $\theta$.  We notice that nonstandard $p$-Brewster angle has a maximum value in terms of $\theta$. This is achieved for ${\theta _M} = \cot {\alpha _i}$, and we see that it is independent of electric and magnetic properties of the system. The value for this the maximum angle can be computed from the expression
\begin{equation}
{\left. {\varphi _B^{\left( s \right)}} \right|_{{\theta _M}}} = {\tan ^{ - 1}}\left\{ {\frac{1}{2}\sqrt {\frac{{{\eta ^2} - {{\csc }^2}{\alpha _i}}}{{{\eta ^2} - 1/{\mu ^2}}}} {{\left[ {{{\left( {1 - 2\sqrt {{\eta ^2} - {{\csc }^2}{\alpha _i}} } \right)}^2} - 4{\eta ^2}} \right]}^{1/2}}} \right\} .
\end{equation}

\section{Discussion}

Results presented here show that electromagnetic radiation propagating in a $\theta$-system bounded in space, and surrounded by a normal medium, exhibits interesting behaviour in polarization properties.  Besides the reported Kerr-Faraday rotation in the polarization plane of the outgoing waves, the Brewster angle is significantly affected by $\theta$ term in boundary conditions for the fields in the surface of the system. A nonstandard $p$-Brewster angle arises, when reflected radiation becomes polarized in the plane of incidence. Compared to the $\theta$-vacuum, we find that magnetoelectric properties properties represented by $\eta$ parameter modify significantly the behaviour. Finally, we  find that the reflected wave could carry a larger part of the energy compared to normal non-$\theta$ systems, which follow Fresnel relations. 

We make a comment on applications to topological insulators,. The quantum approach to those systems implies the periodicity in $2\pi$ of the $\theta$ parameter \cite{Wilczek87}. This is a a consequence of the topological invariance of the Pontryagin term $\int F\wedge F$, which becomes an integer (adequate normalization is needed).\footnote{For normalization, we must implement the substitution $\theta\to \theta e^2/4\pi^2$, $e$ being the adimensional unit charge.} Moreover, for ensuring time reversal (TR) symmetry for TI, the path integral treatment imposes the quantization of $\theta$, being  $0$ or $\pi$ the accepted values, the latter precisely describing TI.  However, this situation which excludes the possibility of considering $\theta$ as a continuous variable, can be circumvented. TR symmetry could be broken in some particular cases leading for instance to spin quantum Hall phenomena; then, different values of $\theta$ are possible in such a case. For  values of $\theta$ within the first zone (Pontryagin number equal to 1), and at first order in $\theta$, from (\ref{tPol}) and (\ref{rPol}) we find that the polarization rotation angle for refracted waves is ${\alpha _t} \sim 2^o$. For the reflected wave, however, a more interesting feature since it becomes dramatically large, ${\alpha _r} \sim {\tan ^{ - 1}}(2/\theta)\sim \pi/2$. Together with these features, already reported in literature, we find that, for small $\theta$, the standard $s$-Brewster angle changes in about $1.2^o$ for $\eta=1.1$. On the other hand, nonstandard $p$-Brewster angle arises at very small values of $\theta$. For $\eta=1.1$ and incident polarization angle of ${\alpha _i} = 10^\circ$, we find that the $p$-Brewster angle is present for approximately $0.02 < \theta  < 11.3$.

The possibility of applying the results here for larger values of $\theta$ is not ruled out in general systems by the identification $\theta \to \theta +2\pi n$, which is the result of considering $\theta$ as an external constant parameter in the quantum theory. Unveiling the dynamics of the scalar field $\phi(x)$ (the "axion" field, in its origin), which could lead the system (not only restricted to TI cases) to collapse in domains, will eventually determine the Pontryagin number resulting for each domain. Some work is in progress \cite{dMHZ}. Therein, and for TR broken symmetry systems, the corresponding value of $\Delta\theta$ between neighboring domains could be obtained. Other possible applications arise in astrophysics and cosmology, where the $\theta$-term could play a role and magnetoelectric properties cannot be neglected. Some work is in progress on these latter issues \cite{dMHZ}.

\noindent ============================
\section*{\Large Acknowledgments}

I am very grateful to J. Zanelli for discussions and suggestions and to Centro de Estudios Cient\'ificos (CECS) for warm hospitality. This work was partially supported by CONICYT-PIA grant ACT-1115.  The P4-Center for Research and Applications in Plasma Physics and Pulsed Power Technology is partially supported by the Comisi\'{o}n Chilena de Energ\'{\i}a Nuclear.   This work is dedicated to my late daughter Mariana Huerta.


\begin{thebibliography}{99}

\bibitem{Chern-Simons} S.~S.~Chern and J.~Simons, Ann. Math. \textbf{99}, 48 (1974).

\bibitem{DJT} S.~Deser, R.~Jackiw and S.~Templeton, Phys. Rev. Lett. \textbf{48}, 975 (1982); Ann. Phys. (NY) \textbf{140}, 372 (1984).

\bibitem{Z-2005} J.~Zanelli, \emph{Lecture notes on Chern-Simons (super-)gravities}, arXiv:hep-th/0502193.

\bibitem{QHE} S.~M.~Girvin and A.~H.~MacDonald, Phys. Rev. Lett. {\bf 58}, 1252 (1987); S.~C. ~Zhang, T.~H.~Hansson and S.~Kivelson, Phys. \ Rev.\ Lett.\ {\bf 62}, 82 (1989); S.~C.~Zhang, Int.\ J.\ Mod.\ Phys.\ {\bf B6}, 25 (1992).

\bibitem{TI2008}X.-L.~Qi, T.~L.~Hughes, and S.~C.~Zhang, Phys. Rev. B {\bf 78}, 195424 (2008), and references therein. 

\bibitem{TI2011}See also X.-L.~Qi and S.~C.~Zhang, Rev. Mod. Phys. {\bf 83}, 1057 (2011) and references therein.


\bibitem{Peccei1977} R.~D.~Peccei and H.~R.~Quinn, Phys. Rev. Lett. {\bf 38}, 1440 (1977).

\bibitem{Wilczek87} F.~Wilczek, Phys.~Rev.~Lett. {\bf 58}, 1799 (1987).

\bibitem{HZ10} L.~Huerta and J.~Zanelli, Proc. Sci., ICFI (2010)
004.

\bibitem{Canfora:2011fd} F.~Canfora, L.~Rosa and J.~Zanelli, Phys. Rev. D {\bf 84}, 105008 (2011).

\bibitem{CS-Grav} M.~Cambiaso, L.~Huerta, A.~Toloza, L.~F.~Urrutia and J.~Zanelli, to be published.

\bibitem{HZ12} L.~Huerta and J.~Zanelli. Phys. Rev. D {\bf 85}, 085024 (2012).

\bibitem{TI-Kerr2010} J.~ Maciejko, X.~L.~Qi, H.~D.~Drew, and S.~C.~Zhang, Phys. Rev. Lett. {\bf 105}, 166803 (2010); W.~K.~Tse and A.~H.~MacDonald, Phys. Rev. B {\bf 82}, 161104 (2010).

\bibitem{Kerr} J.~Kerr, Phil.\ Mag.\ {\bf 3}, 321 (1877); J.~Kerr, Phil.\ Mag.\ {\bf 5}, 161 (1878). See also P.~Weinberger, Phil.\ Mag.\ Lett.\ {\bf 88}, 897 (2008).

\bibitem{dMHZ} F.~de Micheli, L.~Huerta, J.~Zanelli, to be published.

\end{thebibliography}
\end{document}